\documentclass[12pt]{article}
\input{epsf.sty}
\usepackage{times}
\newenvironment{sciabstract}{%
\begin{quote} \bf}
{\end{quote}}

\newcounter{lastnote}
\newenvironment{scilastnote}{%
\setcounter{lastnote}{\value{enumiv}}%
\addtocounter{lastnote}{+1}%
\begin{list}%
{\arabic{lastnote}.}
{\setlength{\leftmargin}{.22in}}
{\setlength{\labelsep}{.5em}}}
{\end{list}}
\title{Kinematic Evidence for an Old Stellar Halo in the Large Magellanic Cloud} 
\author
{Dante Minniti,$^{1\ast}$ Jura Borissova,$^{1}$
 Marina Rejkuba,$^{2}$ \\
David R. Alves,$^{3}$ 
Kem H. Cook,$^{4}$ Kenneth C. Freeman$^{5}$\\
\\
\normalsize{$^{1}$Department of Astronomy, P. Universidad Cat\'olica,}\\
\normalsize{Av. Vicu\~na Mackenna 4860, Casilla 306, Santiago 22, Chile}\\
\normalsize{$^{2}$European Southern Observatory, }\\
\normalsize{Karl-Schwarzschild-Str. 2, D-85748 Garching b.  M\"{u}nchen, Germany}\\
\normalsize{$^{3}$Columbia Astrophysics Laboratory, 550 W. 120th St., New York, NY 10027, USA}\\
\normalsize{$^{4}$IGPP, Lawrence Livermore National Laboratory,}\\
\normalsize{ Livermore, California, USA}\\
\normalsize{$^{5}$Research School of Astronomy and Astrophysics,}\\
\normalsize{ The Australian National University, Mt Stromlo Observatory, Canberra, ACT, Australia}
\\
\normalsize{$^\ast$E-mail:  dante@astro.puc.cl}
}
\date{\today}

\begin{document} 
\maketitle 
\begin{sciabstract}
The oldest and most metal-poor Milky Way stars form a kinematically hot halo, which
motivates the two major formation scenarios for our galaxy:
extended hierarchical accretion \cite{sz78} and rapid collapse \cite{els62}.
RR Lyrae stars are excellent tracers of old and metal-poor populations
\cite{kin91,fea92,ols96}.
We measure the kinematics of 43 RR Lyrae stars 
in the inner regions of the nearby galaxy the Large Magellanic Cloud (LMC).
The velocity dispersion, $\sigma_{true}=53\pm10$ km/s,
indicates that a  kinematically hot
metal-poor old halo also exists in the LMC.
This suggests that our galaxy
and smaller late-type galaxies like the LMC have similar early 
formation histories.
\end{sciabstract}

In the Milky Way, the old metal-poor objects such as globular clusters and 
RR Lyrae stars, define an almost spherical halo population \cite{kin91,fea92,ols96,min96b,
lay98}. Models of halo
formation by accretion \cite{sz78} 
indicate that these old objects formed in small satellite galaxies
which were subsequently accreted by the Galaxy, while dissipational collapse models
\cite{els62}
indicate that the halo formed rapidly before the disk collapsed. If these models apply
to small galaxies, we would expect them
to show a halo population defined by its oldest objects
\cite{min96a}.
At a distance of 50 kpc \cite{alv02}
the ideal laboratory to test this is the LMC, which is 10 times fainter than our Galaxy.
The oldest LMC globular clusters
appear to lie in a flat rotating disk whose velocity
dispersion is 24 km/s \cite{fre83,sch92}. 
This disk 
suggests that the LMC has
indeed no kinematical halo of old metal-poor objects, and that therefore the
formation of the LMC proceeded without a halo phase.
We measured the kinematics of 
field RR Lyrae stars in the LMC which are known to be among the oldest and
most metal-poor objects in this galaxy \cite{ols96}. 
This sample is a by-product of the MACHO microlensing project
\cite{alc97a,alc97b,alc00a}, which 
provided photometry for about
$8000$ RR Lyrae stars in a 10 square degree region around the bar of the
LMC\cite{note1}. 

The observations were acquired with the FORS1 multi-slit
spectrograph at the European Southern Observatory (ESO)
Very Large Telescope (VLT) Unit Telescope 1 (UT1),
during the nights of 10 and 11 January 2003.
Two exposures of 20 minutes were obtained  for each mask
containing 5-10 RR Lyrae stars in six central fields of the LMC bar,
at distances from 0.7 to 1.5 degrees away from the rotational center
(RA = $05^h17^m.6$, DEC = $-69^o02'.0$).
We used the GRIS\_600B+12 grating, that gives $R=1000$ and
covers from $\lambda 3500$ to $\lambda 5900$\AA.
This resolution is adequate for  the measurement of radial velocities
even in the broad-lined RR Lyrae spectra,
provided that a good signal-to-noise is achieved (Fig. 1).
The spectra were reduced using the standard packages APALL and ONEDSPEC
within IRAF.  HeNeAr lamps were used for the wavelength calibration, which
typically have 14 usable lines that yield 0.2 \AA\ rms.
In order to measure the velocities, we used both cross correlations and line
centroiding, and decided that the latter works better for these wide-line
stars, in the presence of line variations with phase. 
Corrections for the phase \cite{smi95}  of the Ca line velocities in a few cases 
do not affect the results.  Tests of both methods 
give similar velocities within 20 km/s rms.

We also measured the kinematical properties of known Mira 
and Cepheid variables from the MACHO and OGLE catalogue. 
We observed 64 RR Lyrae,  5 Cepheids, and 23 Miras of the LMC.
About 10\% of the observing time was devoted to calibrations.
Two masks containing RR Lyrae of the globular cluster $\omega$ Cen
\cite{cle01,may97,kal97} were
acquired using the same setup. Thus, high quality spectra of
17 $\omega$ Cen RR Lyrae (5 RRab, 8 RRc, and 4 RRe)
were obtained.  In addition, a few repeat observations were taken in order to
assess the velocity errors.
The radial velocities (Fig. 2) were measured by centroiding the lines H$\beta$,
H$\gamma$, H$\delta$, and CaII K $\lambda$3933.66 \AA. 
We discarded 21 LMC stars that have only one line measured accurately,
leaving  43 stars that have 2-4 lines accurately measured. The internal
errors measured from the different lines range from 1 to 33 km/s.
Of the final 43 stars considered, 29 are RRab, 11 are RRc and 3 are RRe.
The dispersions for the velocity measurements add quadratically:
$$\sigma^2_{obs} = \sigma^2_{true} + \sigma^2_{rms} + \sigma^2_{phase}$$
where
$\sigma_{obs}$ is the observed velocity dispersion  (Table 1),
$\sigma_{true}$ is the real velocity dispersion of the population,
$\sigma_{rms}$ is the mean error of the individual velocities, and
$\sigma_{phase}$ is the dispersion in the velocities created because
we observe the stars at a random phase.

The $\omega$Cen RR Lyrae velocity dispersion varies as a function of
distance from the cluster center and as a function of
metallicity \cite{may97, nor97}.
For the observed $\omega$Cen field, the velocity dispersion of its RR Lyrae
should be $\sigma_{true}=17$ km/s. We measure $\sigma_{obs}=20$ km/s, indicating
that $(\sigma^2_{rms} + \sigma^2_{phase})^{1/2} = 10$ km/s. The $\omega$Cen 
RR Lyrae spectra have S/N twice as high as the LMC RR Lyrae, and we expect that
$\sigma_{rms ~LMC} > \sigma_{rms ~\omega Cen}$.
The  $\omega$Cen RR Lyrae were also observed at a random phase.
Thus, we expect that the dispersion due to the phase correction of the LMC
RR Lyrae should be low.

The Cepheids can be taken as another control sample for our measurements,
as they have random phase errors comparable to RR Lyrae.  Cepheids 
and carbon stars in the LMC should be kinematically similar.
The velocity dispersion of C stars is $\sigma_{true}=15$ km/s
\cite{alv01,har01,vdm02}.
While we measured only 5 Cepheids, the measured velocity
dispersion $\sigma_{obs}=28$ km/s agrees
with the C stars if
$(\sigma^2_{rms} + \sigma^2_{phase})^{1/2} = 24$ km/s.

The LMC Miras have $\sigma_{true}=33$ km/s \cite{hug91}.
This velocity dispersion is the largest of all kinematic tracers measured
so far in the LMC \cite{gyu00}.
Most of the tracers are of young or intermediate-age,
and show disk kinematics.  In our Galaxy, the Miras have kinematics
intermediate between the disk and the halo \cite{fea92}.
The LMC Miras of our sample yield an observed velocity dispersion
$\sigma_{obs}=43\pm 6$ km/s \cite{note2}.
Subtracting in quadrature 
the true velocity dispersion,
gives $(\sigma^2_{rms} + \sigma^2_{phase})^{1/2} = 28$ km/s,
in agreement with the errors estimated above.

For the final sample of 43 LMC RR Lyrae we measure $\sigma_{obs}=61 \pm 7$ km/s. This
is much larger than the velocity dispersion of any other population, but
in order to find $\sigma_{true}$, we need
to estimate how much is due to errors or the phase correction.
The contribution of phase to the dispersion budget was estimated
to be $\sigma_{phase}=20$ km/s using the radial velocity curves of
known RR Lyrae \cite{cle94, ski93}.
Based on the control samples discussed above, we assume conservatively
$(\sigma^2_{rms} + \sigma^2_{phase})^{1/2} = 30$ km/s.
This yields $\sigma_{true}= 53\pm 10$ km/s for the LMC RR Lyrae \cite{note5}.

One additional correction that cannot be added as a $\sigma$ in quadrature is
the LMC rotation.
Several young and intermediate-age kinematic tracers have been measured in 
the LMC, including HII regions, PN, CH stars, Miras, and carbon stars.
In the inner regions of the LMC bar these populations are rotating
as a solid body, with $25$ km/sec/kpc. For a scale of 1 kpc = 1.2$^o$,
our fields should not show a rotation component larger than 10 km/s.

In addition, a correction for rotation may not be necessary
for the RR Lyrae population, because there is no evidence that this
old population follows
the LMC rotation. Based on the Milky Way RR Lyrae, one might suspect that
the LMC RR Lyrae do not rotate like the rest of the stars.
However, a composite RR Lyrae population may be present. For example, 
earlier interpretation of the RR Lyrae number counts indicated an exponential
disk distribution \cite{alc00c}. Multiple components
(halo + thick disk) cannot be ruled out without rotation measurements.
Our fields are not spread out enough to measure the rotation.
In order to measure the systemic rotation of the RR Lyrae
population, one would need to observe $N\simeq 50$ stars per field in 
fields located $>$3$^o$ away on opposite sides of the bar.
We estimate the correction in two ways: using the velocities from
HI maps \cite{roh84}, and using the mean rotation fits of the 
disk \cite{alv01,vdm02}. 
This correction does not change at all the LMC RR Lyrae velocity dispersion.

The large RR Lyrae velocity dispersion $\sigma_{true}= 53$ km/s
implies that metal-poor old stars are distributed in a halo population. 
The velocity dispersion for the old RR Lyrae stars is higher
than that of the old LMC clusters, 
although there are too few old clusters
to measure the kinematics in the LMC. 
The presence of a kinematically hot, old and metal-poor halo in the LMC
suggests that galaxies like the Milky Way and small galaxies like 
the LMC have similar early formation histories \cite{note3}.

The stellar halo traced by the RR Lyrae 
amounts only to 2\% of the mass of the LMC, which is akin to the Milky Way
halo \cite{kin91,alv01}. In consequence, its
contribution to the microlensing optical depth should not be important \cite{gyu00,alc00b}.
The ongoing Supermacho
experiment would discover an order of magnitude more
microlensing events towards the LMC \cite{stu00}, allowing to test this prediction.

\bibliography{scibib}

\begin{thebibliography}{}
\bibitem{note2} We eliminated a low velocity star,
discrepant by more than 4$\sigma$ 
from the mean, which could be a missidentified foreground star.
\bibitem{note5} We adopt a conservative error of 10 km/s for this
quantity, based on the uncertainties of the control samples.
\bibitem{note3} The sample is not large enough and the velocities are not
accurate enough to (a) detect tidal streams in front or behind the LMC, and (b) 
measure the systemic rotation of the halo RR Lyrae population.
However, from a sample of this size, and with velocities measured to this
accuracy, there is no difficulty in measuring the velocity
dispersion and thus distinguishing a disk population from
a halo population in the LMC.
\end{thebibliography}

\bibliographystyle{Science}

\begin{scilastnote}
\item{We gratefully acknowledge suggestions from E. Olszewski, 
A. Drake, M. Catelan, and C. Alcock,
and support by the Fondap Center for Astrophysics 15010003, by the
U. S. Department of Energy to University of 
California's Lawrence Livermore National Laboratory under Contract 
W-7405-Eng-48, and by the Bilateral Science and Technology Program
of the Australian Department of Industry, Technology and Regional Development.}
\end{scilastnote}

\pagebreak 

\begin{figure}
\begin{center}
\begin{minipage}{17cm}
\epsfxsize=17cm
\epsfbox{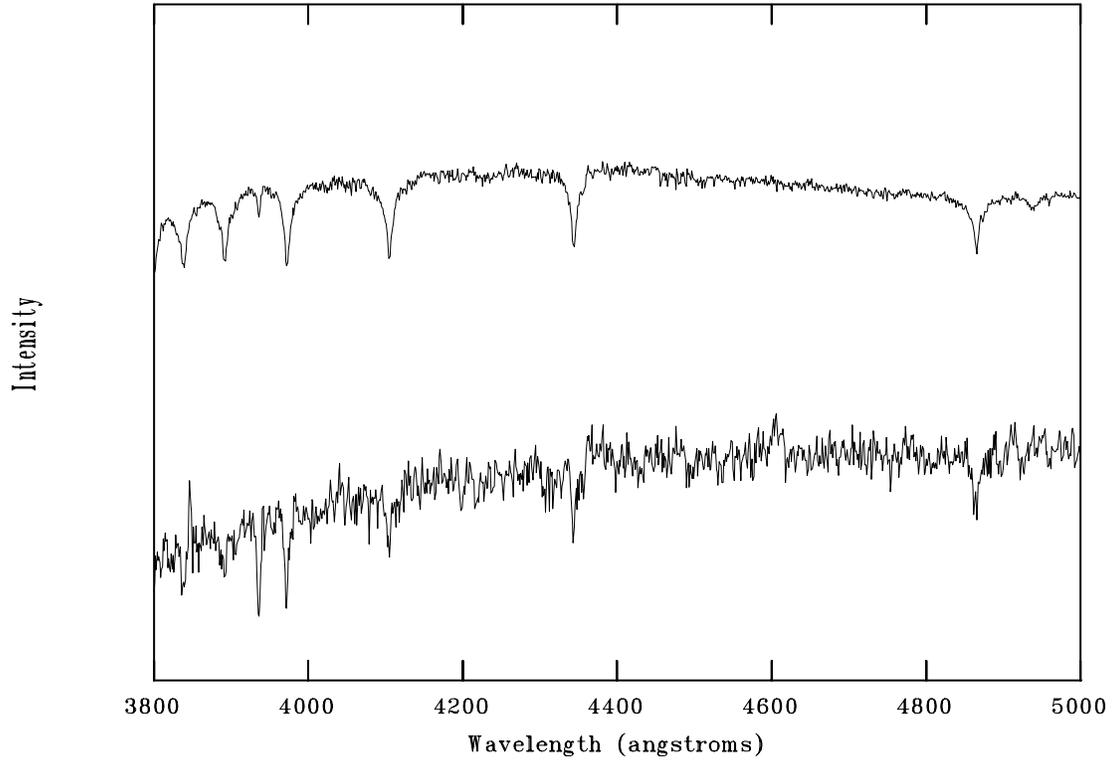}
\end{minipage}
\caption{
Fig. 1: Best spectrum (top) compared with the worst spectrum (bottom)
of the LMC RR Lyrae sample used to measure the LMC kinematics.
}
\end{center}
\end{figure}

\begin{figure}
\begin{center}
\begin{minipage}{15cm}
\epsfxsize=15cm
\epsfbox{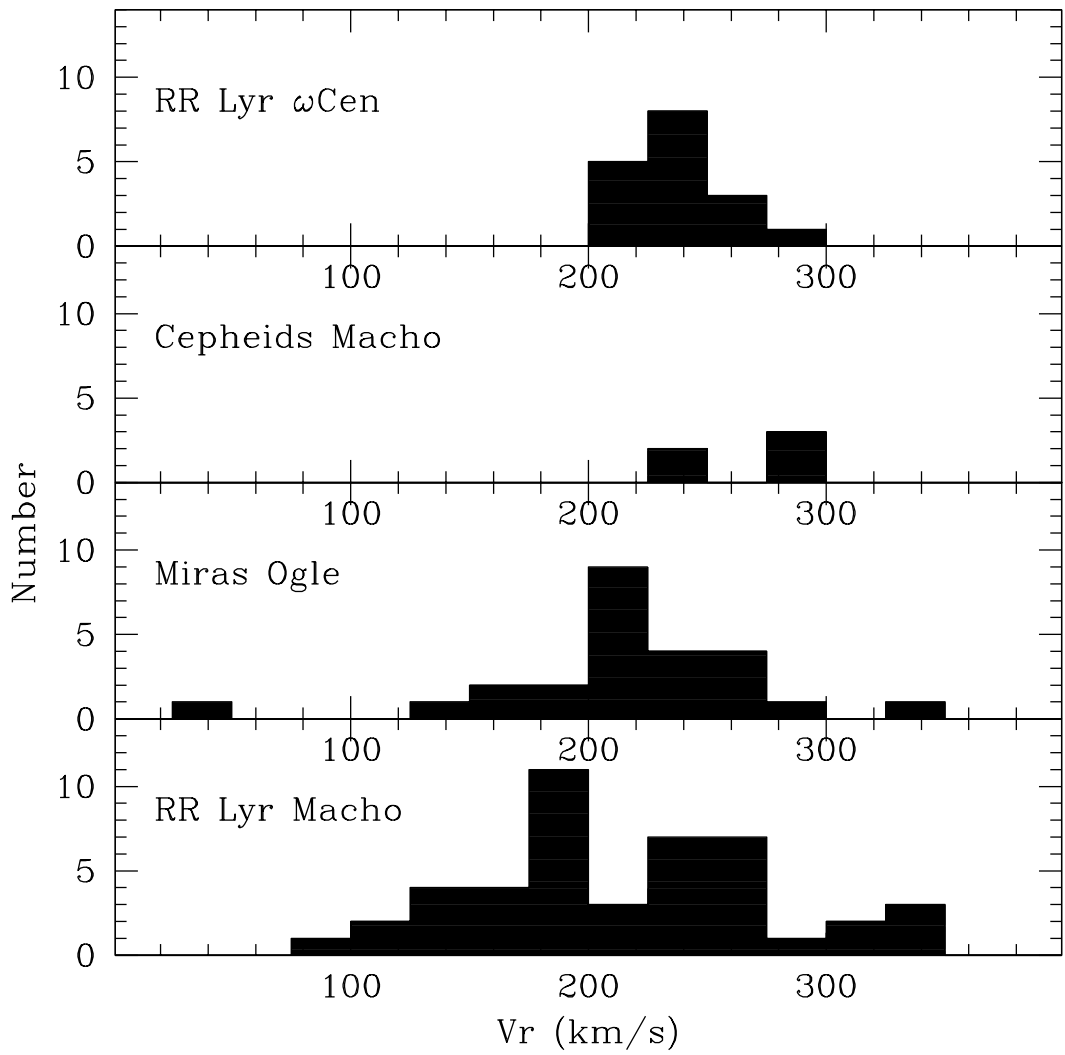}
\end{minipage}
\caption{
Fig. 2: Velocity histograms of LMC RR Lyrae, Cepheids, Miras, and $\omega$ Cen RR Lyrae.
}
\end{center}
\end{figure}

\begin{table}
\caption{Observed Velocity Dispersions}
\label{tab: obs}
\begin{tabular}{ccccc}
\hline
\hline
{Sample}& {N}& {$V$ (mag)}& {$Vr_{mean}$ (km/s)}& {$\sigma_V$ (km/s)}\\
\hline
$\omega$Cen RR Lyrae& 17 &14.0& $237\pm  5$ & $20\pm 3$ \\
MACHO LMC Cepheids  &  5 &16.0& $268\pm 12$ & $28\pm 9$ \\
OGLE LMC Miras      & 23 &17.0& $225\pm  9$ & $43\pm 6$ \\
MACHO LMC RR Lyrae  & 43 &19.5& $214\pm  9$ & $61\pm 7$ \\
\hline
\end{tabular}
\end{table}

\end{document}